# WINTER FORECASTING OF SEPTEMBER/OCTOBER RAINFALL IN THE MURRAY-DARLING BASIN AND SOUTHEASTERN AUSTRALIA

Stjepan Marčelja
Research School of Physics, Australian National University

The atmosphere–ocean system governing Australian climate variability is fundamentally chaotic, characterised by high sensitivity to initial conditions. Major oceanic modes interact over large spatial and temporal scales. In such systems, predictability is not uniform in time but instead depends on the trajectory of the system through its underlying phase space. During austral winter, after the El Niño predictability barrier, sea surface temperature (SST) patterns in the Indian and Pacific Oceans mostly evolve slowly, providing a window of predictability for subsequent spring rainfall. We focus specifically on winter prediction of rainfall during the September/October period, when Indian and Pacific Oceans exert a strong influence over southeastern Australia.

We formulate seasonal rainfall prediction as a reduced-order nonlinear forecasting problem, embedding coupled Indian–Pacific Ocean variability into a low-dimensional state space and projecting it forward using deep neural networks. Variables include Niño 3.4, the Indian Ocean Dipole (IOD), the Indian Ocean meridional SST gradient, and selected empirical orthogonal functions. Monthly time series of the variables then form the input into deep neural networks which project rainfall further into the future.

Forecasts for the 2025 austral spring were generated and archived in the Mendeley database during the winter. Subsequent rainfall data demonstrated a high level of agreement with the forecasts, providing a validation of the method and supporting the hypothesis that chaotic yet conditionally predictable dynamics underpin spring rainfall variability in southeastern Australia.

## 1. Introduction

Seasonal rainfall variability over southeastern Australia exerts a strong influence on water resources, agriculture, and ecosystem functioning, particularly within the Murray–Darling Basin (MDB). Despite decades of research, skilful prediction of austral spring (September/October) rainfall remains challenging, owing to the strong nonlinearity of regional climate drivers and the intermittent nature of large-scale teleconnections (Power et al., 1999; Murphy and Timbal, 2008; Risbey et al., 2009).

A substantial body of work has identified the dominant role of ocean–atmosphere coupling in modulating spring rainfall across southeastern Australia. Variability associated with the El Niño–Southern Oscillation (ENSO) influences the Walker circulation and subtropical jet, with El Niño events typically associated with suppressed spring rainfall and La Niña events favouring wetter conditions (McBride and Nicholls, 1983; Power et al., 1999). IOD further modulates these impacts, with negative IOD events enhancing moisture transport into southern Australia and positive events often reinforcing dry conditions (Saji et al., 1999; Ummenhofer et al., 2009a). Interactions between ENSO and the IOD have been shown to exert a particularly strong influence during austral spring, when both modes tend to reach peak amplitude (Cai et al., 2011; Risbey et al., 2011).

In addition to the IOD, Ummenhofer et al. (2009b) identified an important Indian Ocean index strongly correlated with subsequent spring rainfall over southeastern Australia. This index is based on the meridional SST gradient, defined as the difference between SST anomalies in a southern Indian Ocean region (sI; centred at 30°S, 95°E) and an equatorial eastern Indian Ocean region (eI; centred at 10°S, 110°E). The gradient exhibits a strong relationship with spring rainfall during both dry and wet years.

Numerous statistical and dynamical approaches have been developed to exploit these relationships for seasonal forecasting. Early statistical models employed linear regression or analogue techniques based on SST indices and atmospheric circulation metrics, yielding modest predictive skill at regional scales (Drosdowsky and Chambers, 2001). Subsequent advances incorporated multivariate predictors and empirical orthogonal function (EOF) representations of oceanic variability (Timbal et al., 2010), while coupled dynamical systems such as POAMA and ACCESS-S improved probabilistic skill relative to purely statistical benchmarks (Hudson et al., 2017). Nevertheless, deterministic forecasts of rainfall for specific spring months have generally remained of limited reliability, particularly at lead times extending into austral winter.

Many recent studies have suggested that machine-learning-based methods may capture predictive information that is inaccessible with other methods (Bracco et al, 2025). In our first report (Marčelja, 2025a), we described a method based on information from multiple short time series used as an input to deep learning artificial neural networks trained to predict springtime rainfall in southeast Australia. In the next section we present the theoretical background and more detail on forecasting Spring rainfall based on the reduced complexity of the coupled Indian and Pacific oceans dynamics during the austral winter. By embedding multiple physically motivated indices into a low-dimensional state space and using nonlinear projection, we demonstrate skilful real-time forecasts of September–October rainfall at lead times of 1–3 months.

**2. Ocean dynamics & predictability**

*2.1 Seasonal and decadal modulation of ocean–rainfall teleconnections*

The strength of Indian and Pacific Ocean teleconnections with the rainfall in SE Australia changes throughout a year, over a single decade and over many decades. (Power et al, 1999; Lim et al 2017). Exploring first variations on the intraannual scale, the correlation between ocean variables and rainfall can be evaluated as a monthly table and shown in interpolated images as illustrated in Fig.1.

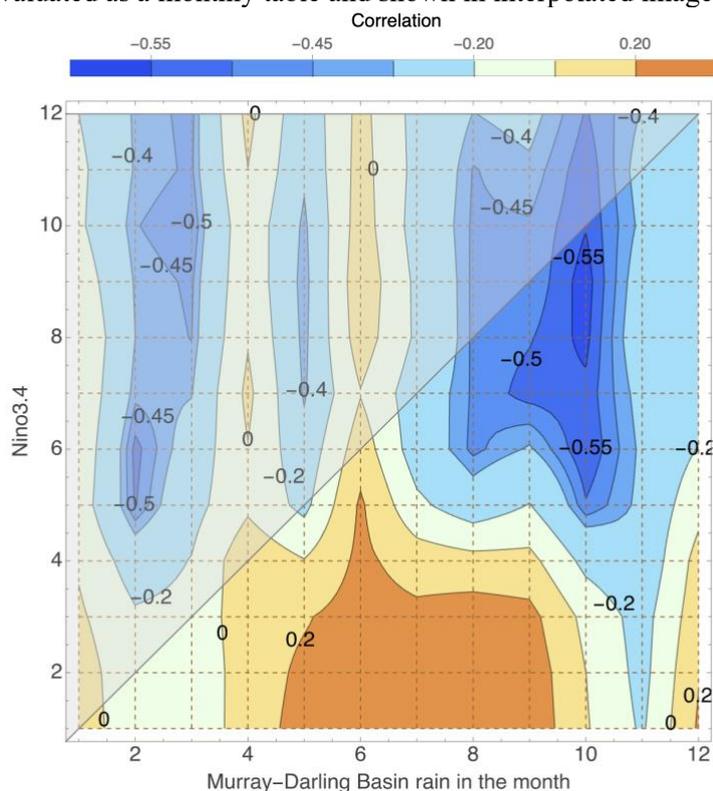

Fig. 1 Correlation coefficient between Niño 3.4 index in any month and MDB rainfall anomaly during another month evaluated over the past 25 years. The lower triangle of the figure shows correlations between the index and future rainfall and therefore contains the information relevant for forecasting.

Such visualisations make it straightforward to identify windows of enhanced predictability. In the example shown, Niño 3.4 July time series are more strongly correlated with September MDB rainfall time series (r = −0.52) than contemporaneous September Niño 3.4 time series (r = −0.46).

On the decadal scale, Pacific decadal oscillations (PDO) are known to influence rainfall on the Australian continent (Power, 1999; Lim 2017). Increased correlation of ENSO and IOD with rainfall coincides with periods of negative PDO index. In the years since cited earlier studies, there was a further increase in correlation between spring rain in southeastern Australia and both IOD and ENSO. In Fig. 2, as an example we show the moving average of the correlation between major Indian Ocean indices and September rainfall in southeastern Australia.

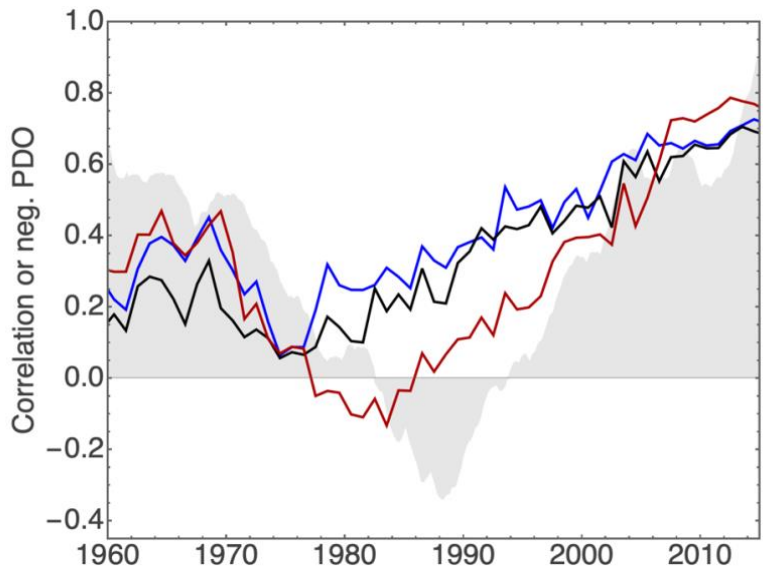

Fig. 2
20-year moving average of the correlation coefficient between southeastern Australia September rainfall and the dominant Indian Ocean EOF (blue), IOD (red) and meridional gradient (black). Also shown in the figure is the PDO index with the reversed sign (grey). The correspondence is not perfect and may change under the conditions of changing climate.

Increasing correlation in the 21[st] century indicates a stronger link between tropical oceans north of the continent and rainfall in southeastern Australia. Another indication is that the occurrence of Australian northwest cloudbands, which bring tropical moisture to southeastern Australia has been increasing over the last 30 years (Reid *et al*, 2019).

Decadal changes in teleconnections (Lim et al 2017) necessitates the restriction of the data used for neural network training to the 21st century. Although this choice shortens the available time series, forecast skill improves markedly when earlier data—drawn from a different climatic regime—are excluded from the training set.

*2.2 Projection from state space to future rainfall*

To prepare for forecasting, the unknown dynamics of the coupled ocean-atmosphere system must be (approximately) embedded into a state space of a given dimension. The state space dimension is the number of variables (or "predictors") to be used in forecasting. If we use *m* time-dependent variables denoted $x_1...x_m$, the state of the dynamical system is described by a time-dependent *m*-dimensional vector $X_m(x_1, x_2, x_3...x_m)$. With a discrete timescale, the projection into the future is a map between the states $X_m(t)$ and future rainfall amounts $R_m(t+T)$, i.e. $R_m(t+T)=f_T(X_m(t))$.
While the real dimension of the ocean and atmosphere dynamics is large, the accuracy of the reconstruction is limited by the length of time series and the presence of noise. We determined the optimal dimension of the reconstruction empirically, where in each case the result was different. For the present task, the often-used method of reconstruction using time delayed values of the same time series is not accurate. The best results are obtained using multivariate embedding (Deyle and

Sugihara, 2011) where we describe the unknown dynamics by selecting the variables with the strongest correlation to the target rainfall.

In the next step we need to find a method of projecting the future rainfall based on the state space reconstruction of the dynamics of the ocean-atmosphere system. In our approach, the transcendental function of $m$ variables $f_T$ is obtained by training a neural network to project from past states to past rainfall amounts. Details of neural network architecture and training methods are described in our earlier report (Marčelja 2025a). A trained neural network represents such a function that for any input vector $X_m(t)$ will output a projected $R_m(t+T)$. These concepts are further illustrated in the next section with a simple 2-dimensional example.

As a test of neural network method performance during the winter, Niño 3.4 forecasts were deposited in the Mendeley database in early July 2025. The results and the comparison with IRI Columbia global models are described in Supplementary material (SM), Sec.1.

*2.3 Two-dimensional representation of coupled Indian Ocean and Pacific Ocean dynamics*

Different aspects of coupling between Indian and Pacific oceans were noticed in both observational and model studies (Kug and Kang, 2006; Zhang et al., 2024; Park et al., 2025). We are mostly interested in the austral winter and spring seasons, when both ENSO and IOD follow relatively stable trajectories.

Tippett et al. (2020) demonstrated that from June to May, the evolution of Niño 3.4 can be captured accurately using only two leading empirical orthogonal functions. Here we demonstrate a major expansion of their finding: During austral winter and spring, *two-dimensional approximation accurately describes the coupled dynamics of the Indian and Pacific Oceans.* Our results reinforce the view that their leading modes form a tightly linked dynamical system.

Using just two modes, ENSO and IOD, entered as Niño 3.4 and Dipole Mode Index (DMI) we reconstruct the dominant dynamics with sufficient fidelity to produce skilful forecasts. As a standard test we hindcast future Niño 3.4 SST with only past Niño 3.4 and IOD time series (SM, Sec. 1). Next, we consider forecasting of MDB rainfall, which proved accurate, even with the approximate two-dimensional state space. This two-dimensional approximation is not assumed to hold universally but applies during austral winter and spring when ENSO and IOD trajectories are relatively stable

Fig. 3a, we illustrate the short SST time series (19 data points each, 6 of which were used for validation) used to train the neural network in 2022. The trained network is then applied to future July conditions to forecast spring rainfall. Results for 2022 and 2023 are hindcasts, while forecasts for 2024 and 2025 (using more data, but very similar to Fig 3) were subsequently confirmed by rainfall data as accurate.

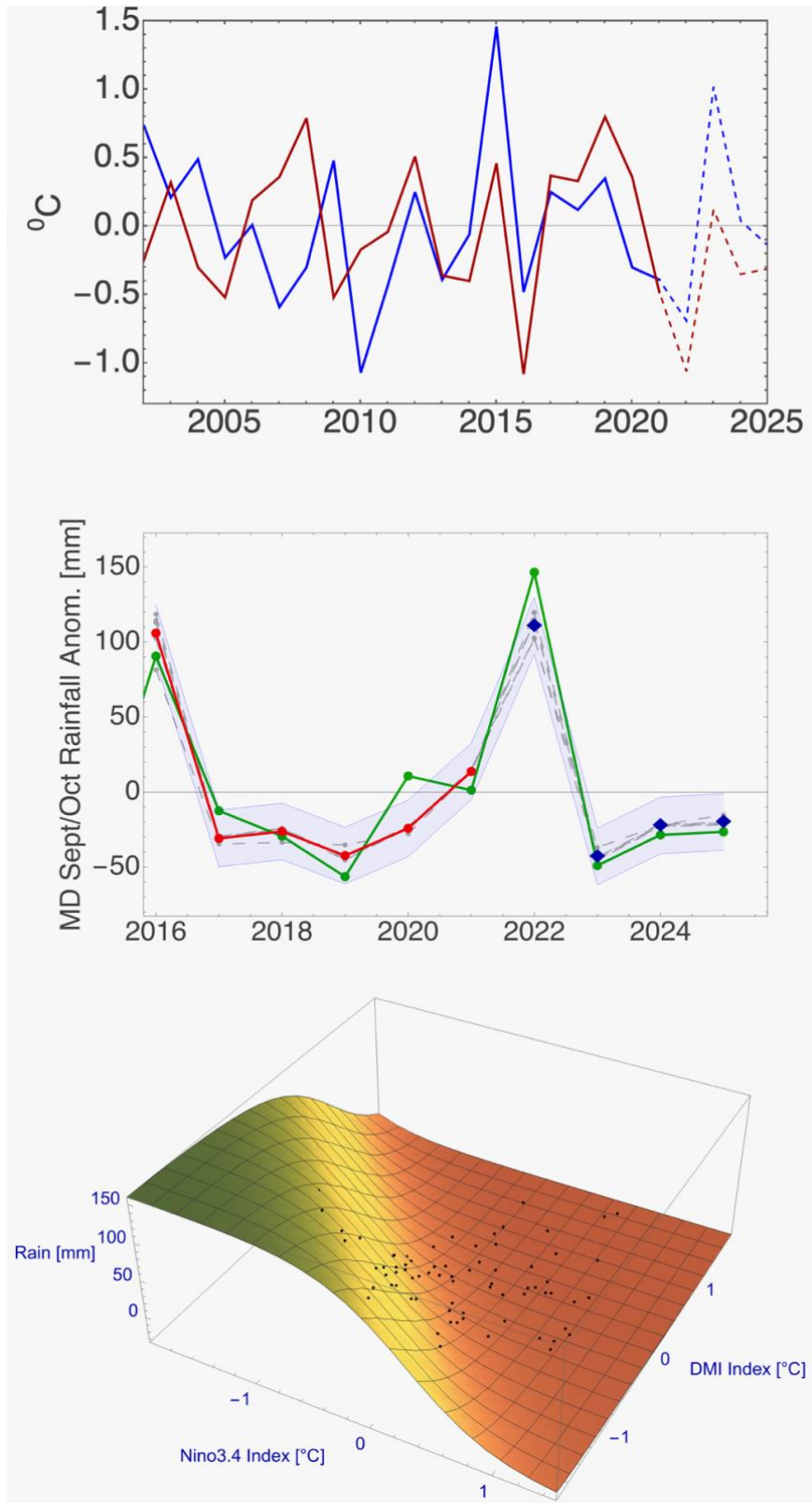

Fig. 3
Top: July Nino 3.4 (blue) and DMI (red) between 2001 and 2025.
Middle: Hindcasting using data from the top panel until 2021. Green: MDB rainfall, red: Neural network validation segment, Diamonds: hindcast of the future rainfall. For actual 2024 and 2025 forecasts we used more data, but since Niño 3.4 and DMI dominate the results were almost the same.
Bottom: Trained neural network from the middle panel envisioned as a nonlinear function of July Niño 3.4 and DMI. Values for all years since 1950 are shown as black points.

*2.4 Linear and nonlinear projections*

With simple two-dimensional state space, it is easy to envision the trained network function $f_T(X_2(t))$ mapping input variables into the future rainfall amount (Fig. 3, bottom). This mapping is inherently nonlinear. Any linear approximation, represented by a plane, fails to capture the curvature imposed by the underlying dynamics. The map outside of the regions with data is an extrapolation and has no significance.

Nonlinear methods, however, are more sensitive to noise and other errors. During periods when low-dimensional state space does not provide a faithful reconstruction of the coupled dynamics, simpler multivariate linear regression will outperform neural networks. This highlights the importance of adapting the forecasting methodology to the prevailing dynamical regime rather than relying on a single technique throughout the year.

**3. Spring 2025 rainfall forecast outcomes**

In the winters of 2024 and 2025, we presented forecasts for combined September/October rainfall for the three regions as defined by the BOM: MDB, Southeastern Australia and Victoria. Forecasts were deposited in independent databases. 2024 forecasts were very close to the measured rainfall, as shown in our earlier work (Marčelja, 2025a).

Our 2025 results are particularly interesting, as we had forecast below average September/October rainfall at the time when major agencies expected a wet spring (SM, Sec. 3). On 16 July 2025 we deposited forecasts based on the June data in the Mendeley database (Marčelja, 2025b). Again, on 5 August 2025 we deposited the forecasts based on July data (Marčelja, 2025c). As a test against a standard benchmark, the second deposition included the prediction of Niño3.4 SST during the 3-months periods SEP/OCT/NOV and OCT/NOV/DEC (described in SM, Sec.3).

The outcomes of predictions based on June and July 2025 data are shown in Table 1 and Fig. 4.

| REGION | DATA MONTH | FORECAST SEPT/OCT RAINFALL (mm) | BOM MEASURED RAINFALL (mm) |
|---|---|---|---|
| Victoria | June | -47 ± 19 | -39 |
|  | July | -55 ± 16 |  |
| MDB | June | -24 ± 22 | -29 |
|  | July | -29 ± 13 |  |
| SE Australia | June | -47 ± 13 | -32 |
|  | July | -44 ± 17 |  |

**Table 1**
Winter forecasts of rainfall for the combined September/October 2025 period. RMS error was evaluated over the validation period. The last column is data for September/October rainfall that were not available at the time when the forecasts shown were deposited.

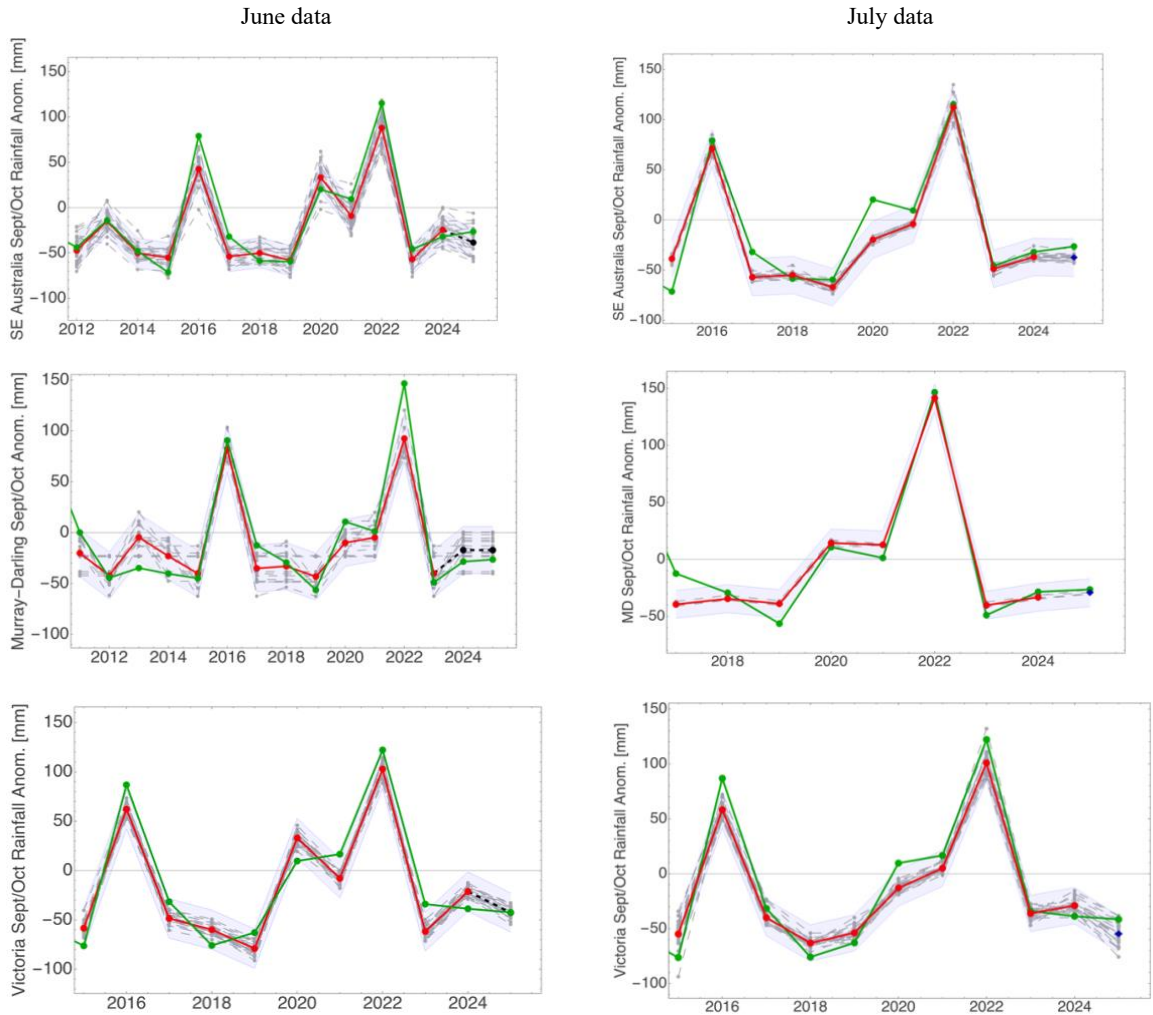

Fig. 4 Winter forecasts (last point) using June data or July data and the corresponding rainfall outcome. Rainfall: green; Neural network validation: red. Different random starts of the network are shown as faint grey lines, and the mean value (shown as diamond) was accepted as the forecast.

The predictions based on June and July data are similar. With July data, the method performs best for the MDB, where rainfall is most influenced by the Indian Ocean. On the other hand, Victoria comprises several different regions, including regions with strong coastal weather influence, resulting in overall lower skill. Overall, forecast skill exceeds climatological and linear regression benchmarks across all three regions, with greatest skill in the MDB.
Once the dynamics of the coupled Indian and Pacific oceans settles into a stable trajectory the window of forecasting possibility remains open until the November rainfall variability ends the temporally limited period of enhanced medium-range predictability.

## 4. Concluding remarks

We presented a successful model of forecasting September and October rainfall in southeastern Australia during the winter months. There are still caveats and limitations that need to be emphasized. The method depends on decadal ocean changes. It would not have been possible 25 years ago and may fade again in the future.
The method was not tested in years of extreme deviations of rainfall from the average. Hindcasting in such years appears to perform well, but until a live test is passed the possibility of author bias cannot be excluded.

The extent of geographical limitations has not been fully tested. In the next stage, more laborious regional forecasting should be attempted, and the geographical coverage may be extended. We began this work by attempting to hindcast for the Mallee region of Victoria with a moderate success (SM, Sec. 2).

Finally, we note that while we had limited success with forecasting based only on past and present data, more successful methods are likely to be built on the synthesis of information from the physical models and deep learning neural network tools (Holgate et al, 2025; Lancelin et al, 2026). While we expect that better systems will soon be developed, the demonstrated real-time skill suggests that reduced-order, regime-aware nonlinear forecasting has practical value for seasonal rainfall prediction in southeastern Australia

SUPPLEMENTARY MATERIAL

1. **Testing Niño 3.4 forecasting**

To test the method against a standard benchmark we deposited on 5 August 2025 into the Mendeley database two Niño 3.4 SST forecasts for Sept/Oct/Nov and Oct/Nov/Dec periods. This format was selected for easy comparison with predictions by 25 dynamical and statistical global models listed monthly by IRI Columbia.

The predictions and the outcomes were as follows:

| Period | Forecast ($^0$C) | Outcome ($^0$C) |
|---|---|---|
| SEP/OCT/NOV | -0.47 | -0.55 |
| OCT/NOV/DEC | -0.61 | -0.62 |

The outcome is illustrated in the next figure

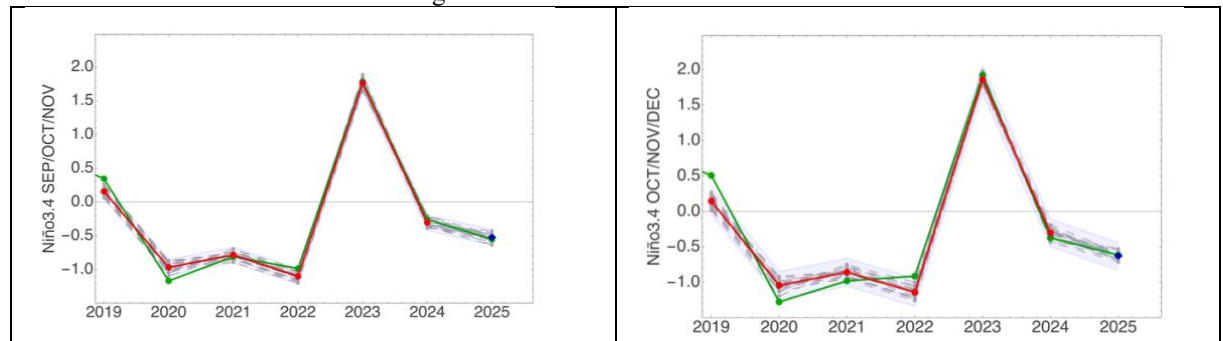

Fig. S1 Niño 3.4 forecast for SON and OND three-month periods evaluated using July data. Niño 3.4 is shown as green, validation data as red and forecast (the last point) as a diamond.

In the tested period, this result places our method equal to the best two of the 25 global models, which predicted Niño 3.4 SST within 0.02 $^0$C of the outcome.

2. **Regional Forecasting: Mallee**

Forecasting rainfall for a specific place is necessarily less accurate. On the timescale of months, the path of local storms is stochastic, and this is reflected in the results. Averaged over larger regions, broader regularities become apparent and forecasting is more accurate.
As an initial test, we attempted to forecast spring rainfall in the Mallee region of Victoria, where advanced knowledge of cropping conditions would be a great advantage. Rainfall was defined as an average of readings from the stations in Mildura, Ouyen, Horsham and Robinvale. As variables describing the dynamical weather conditions, we used only Niño 3.4 and the first two Indian Ocean empirical orthogonal functions. Typical testing for the region appears encouraging, as shown in the following hindcasts for September rainfall evaluated from July data (Fig. S2). Hindcast for the year 2025 was successful, but the very wet year 2022 was only partly anticipated. The results for Mallee shown here are incomplete and preliminary.

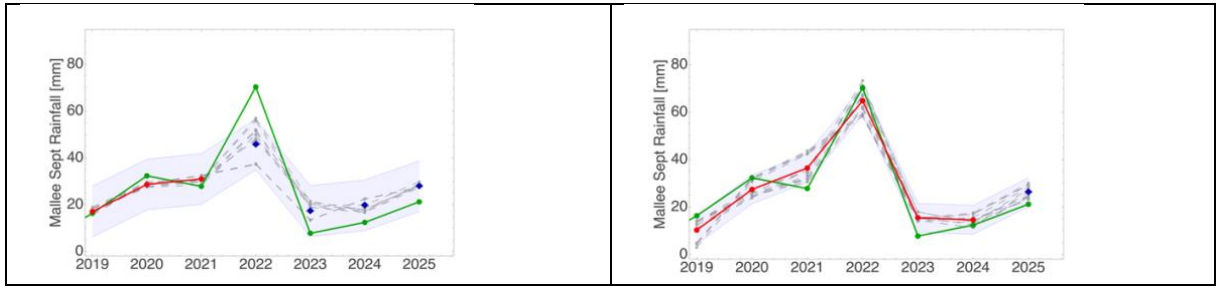

Fig. S2 Hindcasting Mallee September rainfall in 2021 and 2024 using July time series data. Hindcasted values are shown as diamonds.

### 3. Spring 2025 rain outlook by Copernicus Climate Change Service and BOM

EU Copernicus Climate Change Service issues C3S seasonal charts which include outlook for precipitation during future 3-month periods. They are compiled with data from nine global climate agencies.

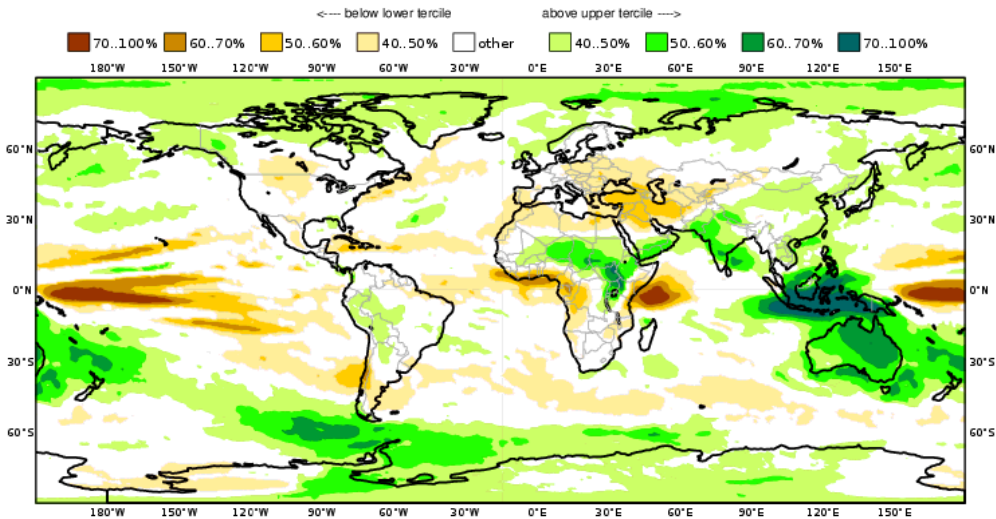
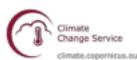

Fig. S3  Copernicus C3S outlook for precipitation during August, September and October 2025 issued on 1/7/2025 (https://climate.copernicus.eu/charts/packages/c3s_seasonal/products/c3s_seasonal_spatial_mm_rain_3m?area=area08&base_time=202507010000&type=tsum&valid_time=202508010000).
In all regions studied in this work, measured August, September and October rainfalls were below average.
BOM forecasts for September and for October issued on 7 August 2025 are available at https://www.bom.gov.au/climate/ahead/outlooks/archive/20250807-outlook.shtml .